\documentclass[sigconf,nonacm]{acmart}
\AtBeginDocument{%
  }

\setcopyright{acmlicensed}
\copyrightyear{2026}
\acmYear{2026}
\acmDOI{10.1145/3858063.3858741}
\acmConference[Conference ICAIF '26]{7th ACM International Conference on AI in Finance}{Nov 14--17, 2026}{Milan, Italy}
\newcommand{\R}{\mathbb{R}}

\usepackage{makecell}
\usepackage{algpseudocode}
\usepackage{comment}
\usepackage{multirow}
\copyrightyear{2026}
\acmYear{2026}
\setcopyright{cc}
\setcctype{by}
\acmConference[ICAIF '26]{7th ACM International Conference on AI in Finance}{November 14--17, 2026}{Milan, Italy}
\acmBooktitle{7th ACM International Conference on AI in Finance (ICAIF '26), November 14--17, 2026, Milan, Italy}
\acmDOI{10.1145/3858063.3858741}
\acmISBN{979-8-4007-3087-0/2026/11}

\begin{document}

\title{Latent Flow Matching for Arbitrage-Aware Implied Volatility Surface Generation}

%\author{anonymous authors}

\author{Dusica Bajalica}
\authornote{Both authors contributed equally to this research.}
\email{dusica.bajalica@dauphine.eu}
\authornotemark[1]
\affiliation{%
 \institution{Université Paris-Dauphine, PSL}
\city{Paris}
 \country{France}
 }

\author{Oscar Brooks}
\authornote{Corresponding author.}
\email{ogb2785@nyu.edu}
\affiliation{%
\institution{Université Paris-Dauphine, PSL \\ New York University, Tandon School of Engineering}
 \city{New York}
 \country{USA}}

\author{Imen Ben Tahar}
%\authornotemark[2]
\email{imen@ceremade.dauphine.fr}

\affiliation{%
 \institution{CEREMADE,   CNRS, Universit\'e  Paris-Dauphine,  \\ PSL  University}
  \city{Paris}
 \country{France}
}

\author{Yating Liu}
%\authornotemark[2]
\email{liu@ceremade.dauphine.fr}
\affiliation{%
\institution{CEREMADE,  CNRS,  Universit\'e  Paris-Dauphine,  \\ PSL  University}
\city{Paris}
\country{France}}

%%\renewcommand{\shortauthors}{Bajalica et al.}


\begin{abstract}
We propose an arbitrage-aware latent flow-matching framework for unconditional implied volatility surface generation. The method first compresses high-dimensional surfaces into a low-dimensional latent space using a variational autoencoder regularized by differentiable calendar-spread,  call-spread and butterfly-arbitrage penalties. A flow-matching model then learns to transport a Gaussian prior toward the empirical latent distribution, and generated latent samples are decoded back into volatility surfaces. We evaluate the approach using marginal and surface-level Wasserstein distances, smile and skew diagnostics, pointwise quantile surfaces, financially interpretable shape metrics, and static no-arbitrage tests. The proposed model closely reproduces the empirical distribution and the main maturity-moneyness structures, achieves the best performance in the extreme $Q_{99}$ regime, and generates $90.8\%$ of surfaces satisfying all tested static no-arbitrage conditions. Overall, the results show that latent flow matching provides a favorable balance between distributional similarity, tail preservation, and financial consistency without requiring post-sampling reweighting.
\end{abstract}

\maketitle

\section{Introduction}

The implied volatility surface (IVS) is a fundamental object in option markets.  It provides a compact representation of information encoded in vanilla option prices across strikes and maturities, including risk-neutral expectations, perceived tail risks,  and stress regimes \cite{gatheral2011volatility}. Accurate modeling of implied volatility surfaces is therefore essential for derivatives pricing, hedging, risk management, and financial scenario generation.

In particular, a smooth and arbitrage-free surface enables the interpolation of prices for unquoted vanilla options and provides a consistent calibration target for local- and stochastic-volatility models used to price and hedge more complex derivatives \cite{dupire1994pricing}. Generative models of implied volatility surfaces can further be used to quantify volatility risk and to generate realistic scenarios for option-portfolio risk measurement and stress testing \cite{cont2002}. 

\subsection{Literature review}
Traditional approaches to volatility surface modeling rely on parametric or semi-parametric representations, such as stochastic volatility models, local volatility models, and SVI-type parameterizations \cite{heston1993, dupire1994pricing, gatheral2014, Fengler2005}. These models are often interpretable and can be designed to satisfy financial constraints, but they may lack the flexibility required to reproduce the full empirical distribution of observed market surfaces. More recently, deep generative models, including GANs \cite{volgan}, VAEs \cite{wang2025controllable}, and diffusion models \cite{jin2025forecasting}, have been proposed to learn volatility surface distributions directly from synthetic or real data. Our work follows this data-driven generative modeling line and introduces latent flow matching as an alternative approach for generating implied volatility surfaces.

It is important to note that, implied volatility surfaces differ fundamentally from ordinary images or generic high-dimensional data. In a financial generation setting, a surface that visually resembles historical observations may still be unusable if it violates static no-arbitrage constraints, since such violations correspond to inconsistent option prices \cite{fengler2009arbitrage, gatheral2014}. This motivates the need for generative models that are both statistically realistic and financially valid. The identification of  model-independent tests for the absence of arbitrage in an option price surface has been addressed in several important works  \cite{laurent2000, carr2005, Buehler2006, Cousot2006, davis-hobson}. An essential  contribution of these works is to formulate necessary and sufficient conditions  to exclude all static arbitrage in terms of  model-independent and numerically tractable  shape constraints on the implied volatility surface.  In this paper, we address the problem of generating implied volatility surfaces under financial consistency requirements. We propose a latent flow matching framework designed to balance two objectives: reproducing the empirical distribution of market surfaces, including average smile and skew profiles as well as tail regimes, while controlling static arbitrage violations. % such as calendar monotonicity and butterfly convexity. 

\subsection{Our contribution}

We propose a two-stage generative framework based on latent flow matching adapted from \cite{lipman2022flow, dao2023flow}. First, we train a variational autoencoder to compress high-dimensional implied volatility surfaces into a low-dimensional latent space. The decoder is regularized by no-arbitrage penalties, encouraging decoded surfaces to lie close to the set of arbitrage-free surfaces.  Second, we train a flow matching model in the latent space, learning a time-dependent vector field that transports a Gaussian prior to the empirical latent distribution. At sampling time, new latent codes are generated by solving the learned ordinary differential equation and then decoded into implied volatility surfaces.

Our main contributions are summarized as follows:

\begin{itemize}
\item We introduce a latent flow matching framework for implied volatility surface generation, reducing the complexity of the generative task by modeling the distribution in a low-dimensional latent space while preserving the structural features of volatility surfaces.

\item We incorporate financial consistency into the generative pipeline through an arbitrage-regularized VAE decoder that accommodates penalties for calendar-spread, call-spread, and butterfly-arbitrage violations.

\item We provide a finance-oriented empirical evaluation against GAN \cite{volgan} and diffusion models \cite{jin2025forecasting}, 
combining distributional metrics with smile and skew diagnostics, tail-quantile analysis, financial-factor distances, and static no-arbitrage compliance tests.

\end{itemize}

Our empirical results show that the proposed latent flow matching model achieves a favorable balance between distributional accuracy and financial validity. In particular, it closely reproduces the marginal implied volatility distribution and average smile/skew profiles, while generating a substantially higher proportion of static no-arbitrage-compliant surfaces than several competing baselines.

\subsection{Organization of the Paper}

The remainder of the paper is organized as follows. Section \ref{sec:method} presents the proposed latent flow matching methodology for implied volatility surface generation. Section \ref{sec:experiment} reports the numerical experiments and empirical evaluation. Section \ref{sec:conclusion} concludes the paper and discusses limitations and future research directions.

% =====================================================
% 2. Method
% =====================================================
\section{Latent Flow Matching for Implied Volatility Surface Generation}\label{sec:method}

This section presents the methodology of the proposed latent flow-matching framework for implied volatility surface generation. Section~\ref{subsec:pipeline} first provides an overview of the two-stage pipeline, and Sections~\ref{subsec:vae} and \ref{subsec:flow-matching} describe the arbitrage-regularized variational autoencoder and the latent flow-matching component in detail.

\subsection{Overview of the Proposed Method}\label{subsec:pipeline}

Let $\{X_n\}_{n=1}^N$ denote a dataset of implied volatility surfaces observed on a fixed grid of maturities and moneyness values. Each surface is represented as
\[
X_n
=
\left(
\sigma_n(\tau_i,m_j)
\right)_{1\le i\le N_\tau,\;1\le j\le N_m}
\in
\mathbb R^{N_\tau\times N_m},
\]%\Yating{on change ici a $\tau_i$ et $m_j$?? et $M$ et $K$ sont $N_{\tau}$ et $N_m$ dans les sections suivantes non ? on uniforme les notations ?? de plus, quelque fois on note $\sigma(\tau, m)$, quelque fois c'est $c(m, \tau)$, quelque fois c'est $w(k, \tau)$ faut uniformer les notations... }
where $\tau_i$ denotes the time to maturity and $m_j$  denotes moneyness. The  pipeline of the proposed method is as follows. In the first stage, a variational autoencoder learns an encoder-decoder pair
\[
E_\phi : X \mapsto q_\phi(z \mid X),
\qquad
D_\theta : z \mapsto \widehat{X},
\]
where $z \in \mathbb{R}^d$ is a low-dimensional latent variable with $d \ll N_\tau N_m$. The decoder is trained not only to reconstruct the input surface, but also to reduce violations of static no-arbitrage constraints. After training, each market surface $X_n$ is encoded into a latent code, typically using the posterior mean
\begin{equation}\label{eq:codes}
    z_n = \mu_\phi(X_n).
\end{equation}
This gives an empirical latent distribution
$
p_{\mathrm{data}}^z
=
\frac{1}{N}
\sum_{n=1}^N \delta_{z_n}$, supported by $\{z_n\}_{n=1}^N$.

In the second stage, we train a flow matching model in the latent space. The model learns a time-dependent vector field
\[
v_\psi : [0,1] \times \mathbb{R}^d \to \mathbb{R}^d,
\]
where $\psi$ denotes the trainable parameters of the neural vector field. The role of $v_\psi$ is to transport a Gaussian prior distribution toward the empirical latent distribution. At sampling time, we draw
$
\widetilde z_0 \sim \mathcal{N}(0,I_d)
$ 
and solve the ordinary differential equation
\[
\frac{d \widetilde z_t}{d t}
=
v_\psi(t,\widetilde z_t),
\qquad
t \in [0,1].
\]
The terminal latent variable $\widetilde z_1$ is then decoded into a generated implied volatility surface $
\widetilde X = D_\theta(\widetilde z_1).$ 
This two-stage construction reduces the dimensionality of the generative problem while preserving the surface structure through the decoder.

\subsection{Arbitrage-Regularized Variational Autoencoder}\label{subsec:vae}

The first component of the model is a variational autoencoder trained to learn a compact representation of implied volatility surfaces. Given an input surface $X$, the encoder outputs the parameters of a diagonal Gaussian posterior
\[
q_\phi(z \mid X)
=
\mathcal{N}
\left(
\mu_\phi(X),
\mathrm{diag}\left(\sigma_\phi^2(X)\right)
\right).
\]
The latent variable is sampled using the reparameterization trick:
\[
z
=
\mu_\phi(X)
+
\sigma_\phi(X) \odot \varepsilon,
\qquad
\varepsilon \sim \mathcal{N}(0,I_d),
\]
where $\odot$ denotes the element-wise (Hadamard) product. Then, the decoder maps $z$ back to a reconstructed volatility surface
\(
\widehat{X} = D_\theta(z).
\)

Recall that the standard VAE objective combines a reconstruction loss and a Kullback-Leibler regularization term 
\begin{equation}\label{eq:normal-vae}
\mathcal{L}_{\mathrm{VAE, standard}}
=
\mathcal{L}_{\mathrm{rec}}
+
\beta \mathcal{L}_{\mathrm{KL}},
\end{equation}
where
\begin{align}
&\mathcal{L}_{\mathrm{rec}}
\!=\!
\mathbb{E}_{q_\phi(z \mid X)}
\left[
\left\| X - D_\theta(z) \right\|_2^2
\right], \mathcal{L}_{\mathrm{KL}}
\!=\!
D_{\mathrm{KL}}
\left(
q_\phi(z \mid X)
\,
\middle\|
\,
\mathcal{N}(0,I_d)
\right),\nonumber
\end{align}
and the parameter $\beta>0$ controls the strength of the latent regularization.
Since the generated objects are implied volatility surfaces rather than generic images, %\Dusica{Ici, je changerai peut-être pour: Since the generated objects are implied volatility surfaces rather than generic images} 
reconstruction accuracy alone is not sufficient. A decoded surface should also be financially meaningful. We therefore augment the VAE objective \eqref{eq:normal-vae} with penalties associated with static no-arbitrage constraints. 

\subsubsection{Static No-Arbitrage Conditions.}
Let $c(\tau,m)=C(\tau,mS)/S$ denote the normalized price of a
European call option, where $m$ is the moneyness and $\tau$
is the time to maturity. Absence of static arbitrage %\Imen{par \cite{davis-hobson} conditions necessaire et suffisante} 
requires normalized call prices
to be non-decreasing in maturity at fixed moneyness, non-increasing in moneyness at fixed maturity, and convex in moneyness at fixed maturity \cite{davis-hobson, Gerhold-Gulum}. Assuming sufficient smoothness, these conditions can be written as
\[
\partial_\tau c(\tau, m)\geq 0,
\qquad
\partial_m c(\tau,m)\leq 0,
\qquad
\partial_{mm}c(\tau, m)\geq 0.
\]
These conditions correspond, respectively, to the absence of calendar-spread, call-spread, and butterfly arbitrage.

\subsubsection{Arbitrage-Penalty Design.}
%\Yating{je suis d'accord, il faut changer les notations} \Dusica{On ne les utilise pas dans le même cas, ie le log-moneyness n'est present dans l'implementation que quand on construit les contraintes de gatheral}
We introduce a calendar arbitrage penalty that penalizes
decreases in the decoded implied volatility surface  across consecutive
maturities.  Denoting by $\widehat{\sigma}(\tau_i,m_j)$ the implied volatility at grid point $(\tau_i,m_j)$ of the decoded surface $\widehat{X}$, we define
\[
\mathcal{L}_{\mathrm{cal}}
=
\frac{1}{N_\tau N_m}
\sum_{i=1}^{N_\tau-1}
\sum_{j=1}^{N_m}
\left[
\widehat{\sigma}(\tau_i,m_j)
-
\widehat{\sigma}(\tau_{i+1},m_j)
\right]_+,
\]
where $[x]_+=\max(x,0)$. 
This penalty encourages implied volatility to be non-decreasing with maturity at each  moneyness. Although stronger than necessary, this condition is sufficient to ensure that the total implied variance
\begin{equation}\label{eq:def-imp-variance}
w(\tau,k)=\tau\widehat{\sigma}^{\,2}(\tau, e^k), \;k=\log m
\end{equation}
is non-decreasing in maturity, thereby satisfying the calendar-spread no-arbitrage criterion of \cite{gatheral2014}.

For butterfly arbitrage, we penalize violations of the Gatheral--Jacquier criterion
\cite{gatheral2014}. For each
fixed maturity $\tau$, define
\begin{align}\label{eq:Gatheral-Jacquier-cond}
g(\tau,k)
:=
&\left(
1-\frac{k\,\partial_k w(\tau,k)}{2w(\tau,k)}
\right)^2
-
\frac{\bigl(\partial_k w(\tau,k)\bigr)^2}{4}
\left(
\frac{1}{w(\tau,k)}+\frac{1}{4}
\right)\nonumber\\
&+
\frac{\partial_{kk}w(\tau,k)}{2},
\end{align}
 where $w(\tau,k)$ is defined by \eqref{eq:def-imp-variance}. 
The absence of butterfly arbitrage requires
$g(\tau,k)\geq 0$.

For the decoded surface, the discrete total
implied variance is defined as
$
w_{i,j}
=
\left(\frac{\widehat{\sigma}_{i,j}}{100}\right)^2\tau_i.
$
Its first- and second-order derivatives with respect to
log-moneyness are approximated by the following central differences with step $\delta k$,
\[
w'_{i,j}
=
\frac{w_{i,j+1}-w_{i,j-1}}{2\delta k},
\qquad
w''_{i,j}
=
\frac{w_{i,j+1}-2w_{i,j}+w_{i,j-1}}{\delta k^2}.
\]
Substituting these quantities into \eqref{eq:Gatheral-Jacquier-cond} yields the
discrete values $g_{i,j}$. We then define
\[
\mathcal{L}_{\mathrm{but}}
=
\frac{1}{N_\tau N_m}
\sum_{i=1}^{N_\tau}
\sum_{j=2}^{N_m-1}
[-g_{i,j}]_{+}.
\]

For the call-spread condition, call prices must be non-increasing in strike. Since moneyness \(m\) is increasing in strike, requiring implied volatility to be non-increasing in \(m\) provides a stronger-than-necessary sufficient condition. We therefore define
\[
\mathcal{L}_{\mathrm{call}}
=
\frac{1}{N_\tau N_m}
\sum_{i=1}^{N_\tau}
\sum_{j=1}^{N_m-1}
\left[
\widehat{\sigma}(\tau_i, m_{j+1})
-
\widehat{\sigma}(\tau_i, m_j)
\right]_+.
\]

\subsubsection{Full arbitrage-regularized VAE loss} Finally, the full arbitrage-regularized VAE loss is
\[
\mathcal{L}_{\mathrm{VAE}}
=
\mathcal{L}_{\mathrm{rec}}
+
\beta \mathcal{L}_{\mathrm{KL}}
+
\lambda_{\mathrm{cal}} \mathcal{L}_{\mathrm{cal}}
+
\lambda_{\mathrm{but}} \mathcal{L}_{\mathrm{but}}
+
\lambda_{\mathrm{call}} \mathcal{L}_{\mathrm{call}},
\]
where $\lambda_{\mathrm{cal}}$, $\lambda_{\mathrm{but}}$  and $\lambda_{\mathrm{call}}$ control the strength of the no-arbitrage regularization. This regularization does not impose exact arbitrage-freeness by construction, but encourages the decoder to map latent variables to surfaces close to the no-arbitrage region.

\subsection{Latent Flow Matching}\label{subsec:flow-matching}

After training the variational autoencoder, we obtain latent codes $\{z_n\}_{n=1}^N$ in \eqref{eq:codes} by encoding the observed market surfaces. The goal of the second stage is to learn a generative model for the empirical latent distribution $p_{\mathrm{data}}^z$. Instead of fitting a generative model directly in the high-dimensional surface space, we apply flow matching in this low-dimensional latent space.

Let $p_0 = \mathcal{N}(0,I_d)$
 be a simple prior distribution, and let
\(
p_1 = p_{\mathrm{data}}^z
\)
be the empirical latent distribution. Flow matching learns a time-dependent vector field
\(
v_\psi : [0,1] \times \mathbb{R}^d \to \mathbb{R}^d
\)
that transports $p_0$ toward $p_1$ along a prescribed family of probability paths.

More precisely, let $z^{0} \sim p_0$ and $z^{1} \sim p_1$ denote two endpoint samples. We consider a smooth interpolation map
\begin{equation}\label{eq:generalIt}
I_t : \mathbb{R}^d \times \mathbb{R}^d \to \mathbb{R}^d,
\qquad
t \in [0,1],
\end{equation}
satisfying
\(
I_0(z^{0},z^{1}) = z^{0}, 
I_1(z^{0},z^{1}) = z^{1}.
\)
The intermediate latent state is defined by
\(
z_t = I_t(z^{0},z^{1}),
\)
and the corresponding target velocity is
\[
u_t(z^{0},z^{1})
=
\frac{d}{dt} I_t(z^{0},z^{1}).
\]
The flow matching loss is
\[
\mathcal{L}_{\mathrm{FM}}(\psi)
=
\mathbb{E}_{t,z^{0},z^{1}}
\left[
\left\|
v_\psi(t,z_t)
-
u_t(z^{0},z^{1})
\right\|_2^2
\right],
\]
where $t$ is sampled uniformly from $[0,1]$, $z^{0} \sim p_0$, and $z^{1} \sim p_1$. This objective trains the neural vector field to approximate the velocity associated with the chosen path between the prior and the empirical latent distribution.

This formulation of $I_t$ in \eqref{eq:generalIt} is general and allows different choices of interpolation paths. For instance, the standard linear path is
\(
I_t^{\mathrm{lin}}(z^{0},z^{1})
=
(1-t)z^{0} + t z^{1},
\)
with target velocity
\[
u_t^{\mathrm{lin}}(z^{0},z^{1})
=
z^{1} - z^{0}.
\]
Another possible choice is the trigonometric path
\begin{equation}\label{eq:trigonometric}
I_t^{\mathrm{trig}}(z^{0},z^{1})
=
\cos\left(\frac{\pi t}{2}\right) z^{0}
+
\sin\left(\frac{\pi t}{2}\right) z^{1},
\end{equation}
whose target velocity is
\begin{equation}\label{eq:velocity-trigonometric}
u_t^{\mathrm{trig}}(z^{0},z^{1})
=
-
\frac{\pi}{2}
\sin\left(\frac{\pi t}{2}\right) z^{0}
+
\frac{\pi}{2}
\cos\left(\frac{\pi t}{2}\right) z^{1}.
\end{equation}

Once the vector field has been learned, new latent samples are generated by solving the ordinary differential equation
\[
\frac{d \widetilde z_t}{d t}
=
v_\psi(t,\widetilde z_t),
\qquad
\widetilde z_0 \sim \mathcal{N}(0,I_d),
\]
from $t=0$ to $t=1$. The terminal point $\widetilde z_1$ is then passed through the decoder 
\(
\widetilde X = D_\theta(\widetilde z_1).
\)
The resulting sample $\widetilde X$ is a generated implied volatility surface.

\textbf{Remark.} The advantage of this latent formulation is twofold. First, it avoids modeling the full high-dimensional surface distribution directly. Second, the decoder learned in the first stage acts as a structural map from latent variables to volatility surfaces, allowing the flow model to focus on learning the distribution of market-relevant latent factors rather than pointwise surface variations.

% =====================================================
% 4. Experiments
% =====================================================
\section{Experiments}\label{sec:experiment}

This section presents the empirical evaluation of the proposed framework. Section~\ref{subsec:data} describes the dataset and the construction of the implied volatility surfaces. Section~\ref{subsec:training} provides the architecture and training details, while Section~\ref{subsec:benchmarks} introduces the benchmark methods. Section~\ref{subsec:metrics} defines the statistical and financial evaluation criteria, and Section~\ref{subsec:results} presents and discusses the numerical results. Finally, Section~\ref{subsec:Ablation} reports ablation and sensitivity analyses examining the VAE loss design, the distributional cost of nearly arbitrage-free generation, and the choice of latent-space dimension.

The implementation is available at \href{https://github.com/DusBaja/ivs-generative-benchmark}{https://github.com/DusBaja/ivs-generative-benchmark}. %The repository also contains additional exploratory comparisons with Schr\"odinger bridge-based generative methods for implied volatility surfaces.

\subsection{Data and Implied Volatility Surface Construction}\label{subsec:data}Implied volatilities are extracted from end-of-day SPX option data obtained from OptionsDX over the period from January 2020 to December 2023.  The resulting training set comprises 1,000 implied volatility surfaces. For each trading date, options with maturities between 7 and 365 calendar days and moneyness in the interval [0.8,1.2] are retained.  Following standard market practice, out-of-the-money put implied volatilities are used for strikes below the at-the-money level, whereas out-of-the-money call implied volatilities are used for strikes above it for better liquidity. Implied volatilities are observed for a discrete set of expiries and strikes. The grid of available observations is typically irregular and varies across trading dates. In order to evaluate each surface on the same fixed grid comprising 32 moneyness levels and 16 maturities, an SVI parametrisation is calibrated and interpolated %fitted at each maturity 
following Gatheral and Jacquier~\cite{gatheral2014}. The SVI parametrisation is used to obtain a smooth representation of each volatility smile on the common moneyness grid despite the irregular and date-dependent location of the observed strikes. Interpolation across maturities is then performed in total variance. Any remaining missing values are filled using nearest-neighbour interpolation, after which calendar-arbitrage corrections are applied by enforcing total implied variance to be non-decreasing with maturity through a cumulative-maximum adjustment.

The resulting surface is flattened into a vector in \(\R^{N_mN_{\tau}}=\R^{32\times 16}\). Surfaces are normalised to $[-1, 1]$ using the 4th and 96th percentiles of the training distribution as bounds.

\subsection{Architecture and Training Details}\label{subsec:training}

\paragraph{Arbitrage-Regularized VAE}
The encoder maps each flattened implied volatility surface
$x \in \R^{32\times 16}= \mathbb{R}^{512}$ to a diagonal Gaussian posterior $q_\phi(z\mid x)
=
\mathcal{N}\big(
\mu_\phi(x),
\operatorname{diag}\big(\sigma_\phi^2(x)\big)
\big)$ on $\R^6$.
Both the encoder and decoder are implemented as residual multilayer perceptrons with skip connections. The encoder uses hidden widths $(256,128,64)$, while the decoder mirrors this architecture with widths $(64,128,256)$.  Each residual block consists of two linear layers with a ReLU activation and a linear shortcut. The decoder ends with a linear projection onto $\mathbb{R}^{512}$, yielding a reconstructed surface $\widehat{x}$ in the normalized data space. After training, the VAE parameters are frozen, and the posterior means of the encoded surfaces are used as samples from the empirical latent distribution for training the flow-matching model.

\paragraph{Latent Flow Matching.}
The latent vector field
$v_\psi:\mathbb{R}^{6}\times[0,1]\rightarrow\mathbb{R}^{6}$
is parameterized by a U-Net-style residual MLP with hidden widths
$(64,128,64)$. Skip connections are used throughout the network,
and time information is incorporated into each block through
adaptive layer normalization. Given a Gaussian sample
$z_0\sim\mathcal{N}(0,I_6)$ and an empirical latent code $z_1$, we
use the trigonometric interpolation \eqref{eq:trigonometric} and \eqref{eq:velocity-trigonometric}.
The vector field is trained to approximate the target velocity
$u_t^{\mathrm{trig}}$. At generation time, the learned ODE is integrated from
$t=0$ to $t=1$ using 100 Euler steps. 

Table~\ref{tab:hyperparams} summarizes the architecture,
optimization, and sampling hyperparameters selected after tuning
in the two-stage generative framework.

\begin{table}[!t]
\centering
\caption{Hyperparameters for VAE and Flow Matching.}
\label{tab:hyperparams}
\begin{tabular}{llc}
\toprule
Component & Hyperparameters & Value \\
\midrule
\multirow{11}{*}{VAE}
 & Latent dimension $d$            & 6 \\
 & Encoder widths                  & $(256, 128, 64)$ \\
 & Decoder widths                  & $(64, 128, 256)$ \\
 & KL weight $\beta$               & $10^{-2}$ \\
 & Calendar weight $\lambda_\text{cal}$ & $3 \times 10^{-2}$ \\
 & Butterfly weight $\lambda_\text{but}$ & $2 \times 10^{-3}$ \\
  & Call weight $\lambda_\text{call}$ & 0 \\
 & Learning rate                   & $2 \times 10^{-3}$ \\
 & Batch size                      & 64 \\
 & Epochs                          & 1200 \\
 & KL warm-up epochs                & 200 \\
\midrule
Flow & Path                            & Trigonometric  \\
\multirow{6}{*}{Matching} & U-Net hidden widths              & $(64, 128, 64)$ \\
 & Time embedding dim.        & 32 \\
 & Learning rate                   & $10^{-4}$ \\
 & Batch size                      & 64 \\
 & Epochs                          & 1200 \\
 & Euler steps at sampling           & 100 \\
%\midrule
%\multirow{Data}
% & Grid                            & $16 \times 32$ \\
% & Percentile scaling bounds                    & $(q_{0.04},\, q_{0.96})$  \\
% & Training surface number               & 1000 \\
\bottomrule
\end{tabular}
\end{table}

\subsection{Benchmark Methods}\label{subsec:benchmarks}

We compare the proposed latent flow matching model, denoted by \textbf{L-FM}, with several recent deep generative baselines for implied volatility surface generation. These include the unconditional adaptations of a score-based diffusion model trained with an arbitrage penalty \cite{jin2025forecasting}, as well as the raw and reweighted variants of VolGAN \cite{volgan}. VolGAN Raw is trained to reproduce the empirical distribution of volatility surfaces without any post-sampling adjustment, whereas the reweighted variant applies an additional correction step to improve compliance with static no-arbitrage constraints. For simplicity, we refer to the reweighted variant as VolGAN throughout the remainder of the paper.

\subsection{Evaluation Metrics}\label{subsec:metrics}

We evaluate the generated surfaces from both statistical and financial
perspectives.

\paragraph{Distributional Similarity.} Marginal distributional similarity is measured by the Wasserstein-1 distance $\mathcal{W}_1$ between implied volatility values pooled across all surfaces and grid points. To account for the joint structure of the full surfaces, each surface is flattened into a vector in $\mathbb{R}^{512}$ and compared using the sliced Wasserstein distance (SWD) over 300 random projections. Lower distance values indicate better agreement with the empirical distribution.

\paragraph{Shape and Tail} We further compare the average volatility smile and skew at representative short, medium, and long maturities. Tail behavior is assessed through pointwise quantile surfaces at levels $q\in\{0.05,0.25,0.50,0.75,0.95,0.99\}$.

\paragraph{Financial Metrics Diagnostics.}%\Imen{Je propose le titre : Financial metrics Diagnostics}
We construct four metrics  $\mathfrak{L}, \mathfrak{T},  \mathfrak{S}, \mathfrak{C}$ to capture financially significant shape components of the implied volatility surface, which are: the level, the term structure,  the skew and the curvature. For a volatility surface \(\widehat{\sigma}\in\mathbb{R}^{N\tau\times N_m}\), let \(m^{*}\) denote the at-the-money moneyness level, \(\mathcal{I}_S\) and \(\mathcal{I}_L\) the shortest- and longest-maturity quartiles, and \(\mathcal{I}_P\) and \(\mathcal{I}_C\) the put- and call-wing moneyness quartiles, and define 
\begin{align}\label{eq:def-financial-factors}
& \mathfrak{L}(\widehat{\sigma}) = \frac{1}{N_\tau}\sum_{i=1}^{N_\tau}
    \widehat{\sigma}(\tau_i, m^*), \nonumber\\
&\mathfrak{T}(\widehat{\sigma}) =
    \frac{1}{|\mathcal{I}_L|}\sum_{i \in \mathcal{I}_L}
        \widehat{\sigma}(\tau_i, m^*)
    -\frac{1}{|\mathcal{I}_S|}\sum_{i \in \mathcal{I}_S}
        \widehat{\sigma}(\tau_i, m^*), \nonumber\\
&\mathfrak S( \widehat{\sigma}) =
    \frac{1}{|\mathcal{I}_P|N_\tau}
    \sum_{i,j\,:\,j\in\mathcal{I}_P}\widehat{\sigma}(\tau_i,m_j)
    -\frac{1}{|\mathcal{I}_C|N_\tau}
    \sum_{i,j\,:\,j\in\mathcal{I}_C}\widehat{\sigma}(\tau_i,m_j), \nonumber\\
&\mathfrak C( \widehat{\sigma}) =
    \frac{1}{2}\Bigl(
        \overline{\sigma}_P + \overline{\sigma}_C
    \Bigr) - \mathfrak{L}(\widehat{\sigma}) , 
\end{align}
where \(\overline{\sigma}_P\) and \(\overline{\sigma}_C\) are the mean implied volatilities over the put and call wings. %\Yating{dans la def de $\mathfrak C( \widehat{\sigma})$, a la fin c'est $- \mathfrak{L}(\widehat{\sigma})$ a la place de ``- level'' ? }%\Yating{$\overline{\sigma}_P$ et $\overline{\sigma}_C$ non defini}
%*** equations***
We compute the \(\mathcal{W}_1\) distances between the distributions of the four financially interpretable  metrics $\mathfrak{L}, \mathfrak{T}, \mathfrak{S}, \mathfrak{C}$ evaluated on the training and generated datasets.

\paragraph{No-Arbitrage Validity.} Finally, static no-arbitrage validity is measured as the percentage of generated surfaces satisfying the calendar-spread, call-spread, and butterfly inequalities over the entire grid. Higher validity is better.  

Considering these metrics jointly allows us to assess the trade-off between distributional fidelity and financial consistency.

\subsection{Numerical Results and Discussion}\label{subsec:results}

We compare L-FM with the arbitrage-regularized diffusion model of \cite{jin2025forecasting}, VolGAN Raw and VolGAN from \cite{volgan}. %{\color{red}, and three Schr\"odinger bridge baselines: SBJTS-PCA, SBJTS-AE/VAE, and LightSB.}  
%We generate $5{,}000$ surfaces from each model with over 5 independent training runs.
For each model, we perform five independent runs and generate 5,000 surfaces per run.  Throughout the tables, the best-performing result for each metric is shown in \textbf{bold}, while the second-best result is \underline{underlined}.

% ============================================================
\subsubsection{Distributional Similarity.}

We first compare the marginal implied volatility distributions obtained by pooling values across all generated surfaces, maturities, and moneyness points. 

\begin{figure}
    \centering
    \includegraphics[width=1\linewidth]{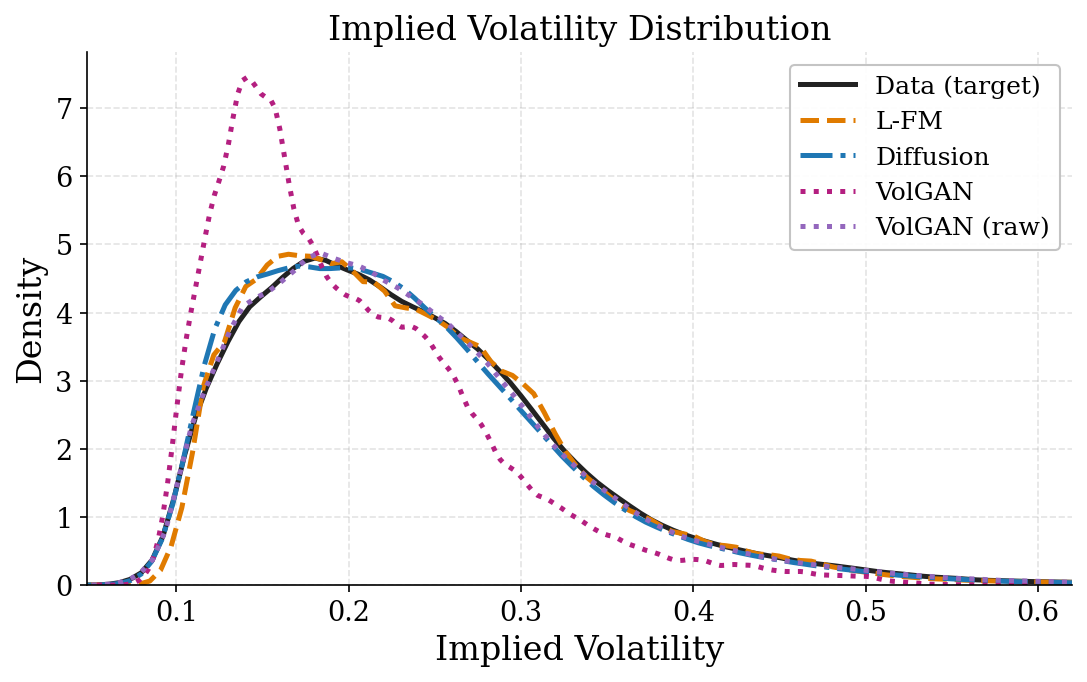}
\caption{Marginal implied volatility distributions}
    \label{fig:distribution_diagnostics}
\end{figure}

Figure~\ref{fig:distribution_diagnostics} shows that L-FM closely reproduces both the central mass and the right tail of the empirical distribution. Diffusion and VolGAN Raw also track the empirical distribution reasonably well, whereas VolGAN places excessive probability mass at low volatility levels and substantially underrepresents the right tail.

\begin{figure}
    \centering
    \includegraphics[width=0.95\linewidth]{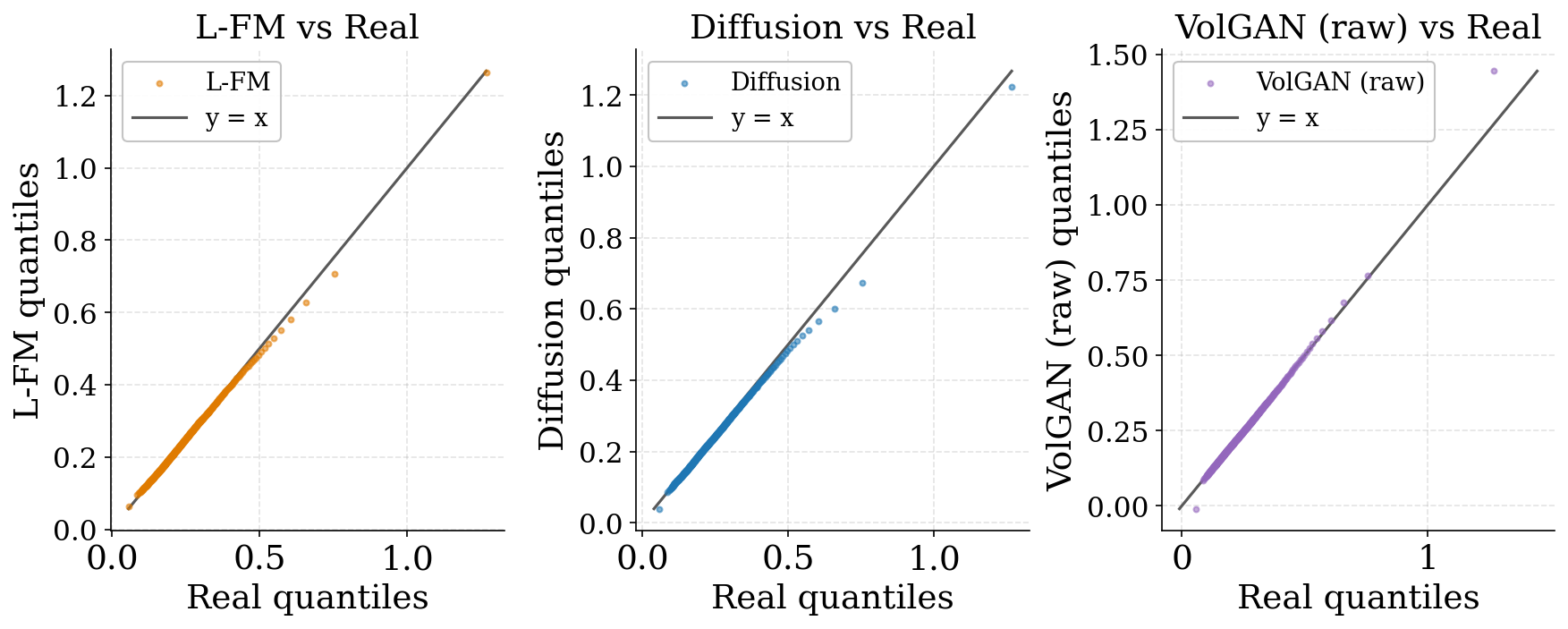}
    \caption{Q--Q plots of generated versus empirical marginal implied volatility distributions.}
    \label{fig:qq}
\end{figure}

The Q--Q plots in Figure~\ref{fig:qq} provide a quantile-wise comparison.  L-FM closely matches the empirical quantiles across most of the distribution, with only slight deviations at the extremes. Diffusion and VolGAN Raw also perform well overall, but exhibit larger discrepancies in the upper tail.

%L-FM remains very close to the diagonal across nearly the entire distribution, indicating an accurate match to the empirical quantiles. Diffusion also tracks the diagonal well over the bulk of the distribution, but departs from it in the upper tail, suggesting an underrepresentation of the largest implied volatility values. VolGAN Raw stays close to the diagonal for most quantiles as well, although it shows more noticeable discrepancies at the most extreme quantiles, particularly at the upper end.

%The Q--Q plots in Figure~\ref{fig:qq} provide a quantile-wise comparison. L-FM remains close to the diagonal over most of the distribution, with only a moderate deviation at the most extreme observations. By contrast, the deviations of Diffusion and VolGAN become increasingly pronounced in the upper tail, indicating an underrepresentation of the largest implied volatility values.

%\Yating{``Perfect fit'' dans figure \ref{fig:qq} est un peu bizarre ... et cette expression est pas tres academic...je propose qu'on utlise : the identity line $y=x$}

Table~\ref{tab:wasserstein} reports the global Wasserstein-1
distance $\mathcal{W}_1$ between the pooled implied volatility
distributions and the sliced Wasserstein distance (SWD) between the flattened $512$-dimensional surface distributions. L-FM achieves the smallest global \(\mathcal{W}_1\) and the second-smallest SWD, while VolGAN Raw ranks second in global \(\mathcal{W}_1\) and first in SWD. Both substantially outperform Diffusion and VolGAN.

\begin{table}[ht]
\centering
\caption{Global $\mathcal{W}_1$ distance and sliced Wasserstein
distance against the empirical surface distribution. Lower is
better.}
\label{tab:wasserstein}
%\small
\begin{tabular}{lcc}
\toprule
Model & Global $\mathcal{W}_1$ & SWD \\
\midrule
L-FM              &  \textbf{0.00283} $\pm$ 0.00144 & \underline{0.00697} $\pm$ 0.00056 \\
Diffusion         & $0.00694 \pm 0.00283$          & $0.00782 \pm 0.00180$ \\
VolGAN  & $0.03447 \pm 0.00607$          & $0.02960 \pm 0.00426$ \\
VolGAN Raw        & \underline{0.00287} $\pm$ 0.00113          & \textbf{0.00491} $\pm$ 0.00078 \\
\bottomrule
\end{tabular}
\end{table}

\subsubsection{Shape and Tail}

\begin{figure*}[ht]
    \centering
    \includegraphics[width=0.9\linewidth]{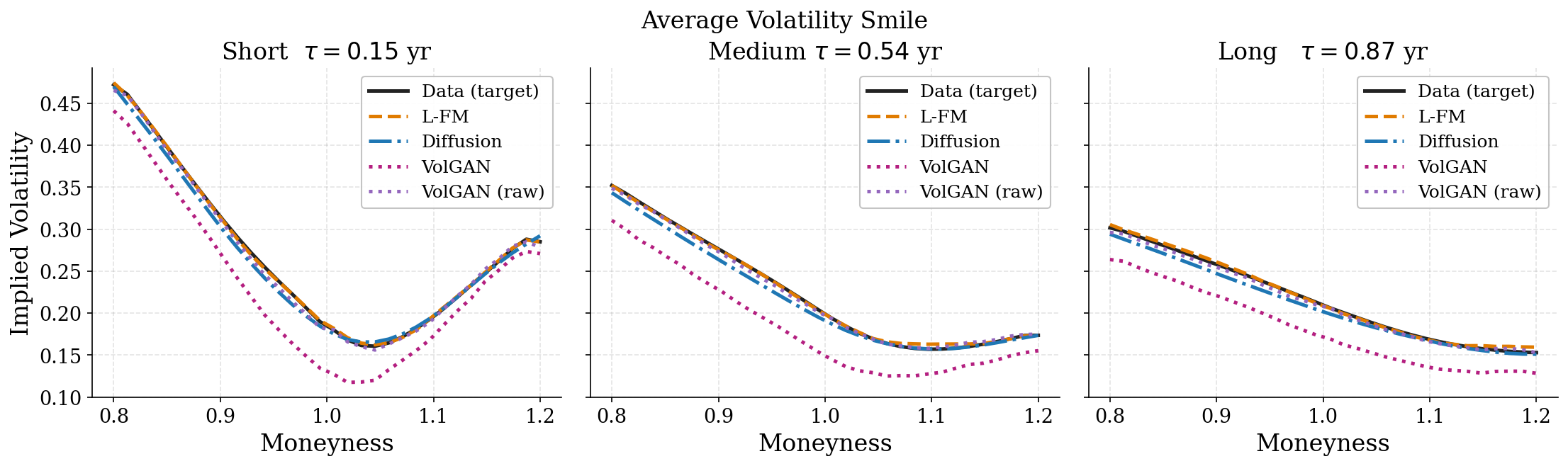}
 %   \vspace{0.2cm}
    \includegraphics[width=0.9\linewidth]{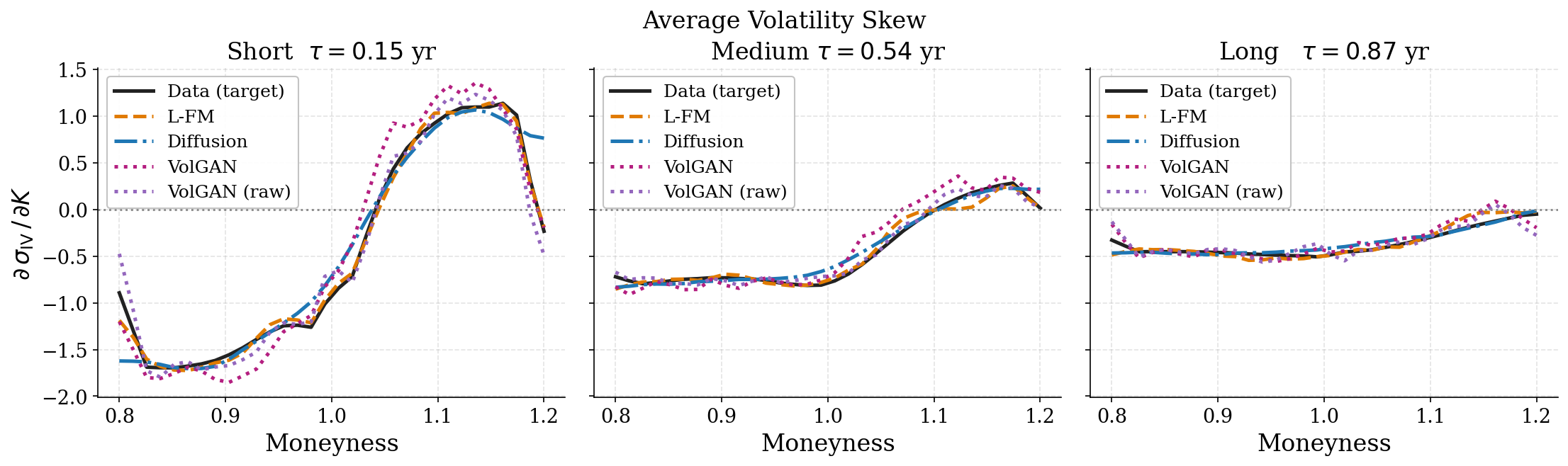}
    \caption{Average implied volatility smiles and numerical skews at short, medium, and long maturities.}
    \label{fig:smile_skew}
\end{figure*}

Figure~\ref{fig:smile_skew} compares the average implied volatility smile and numerical finite-difference skew at representative short, medium, and long maturities. L-FM closely follows the empirical smile profiles and captures the steep short-maturity put wing together with the progressive flattening of the smile as maturity increases. VolGAN Raw also remains close to the data, whereas Diffusion moderately underestimates implied volatility over several moneyness regions. VolGAN exhibits a more pronounced downward shift, particularly at medium and long maturities.

The empirical numerical skew is strongly negative on the put wing, then increases toward zero and becomes positive over part of the call wing at short and medium maturities. This transition becomes flatter at longer maturities. L-FM reproduces the main shape and maturity dependence of the skew, although local discrepancies remain near the boundaries of the moneyness grid.

\begin{figure*}[ht]
    \centering
    \includegraphics[width=0.99\textwidth]{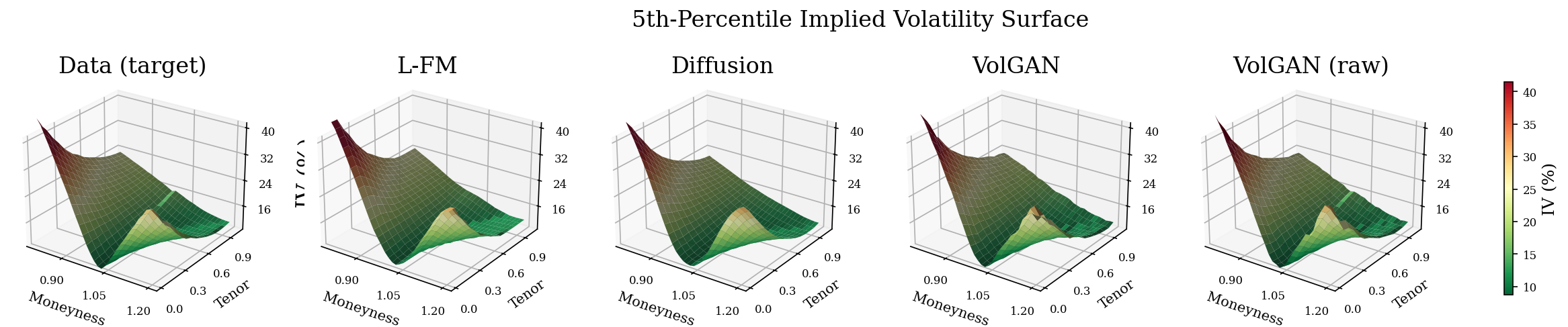}

    \includegraphics[width=0.99\textwidth]{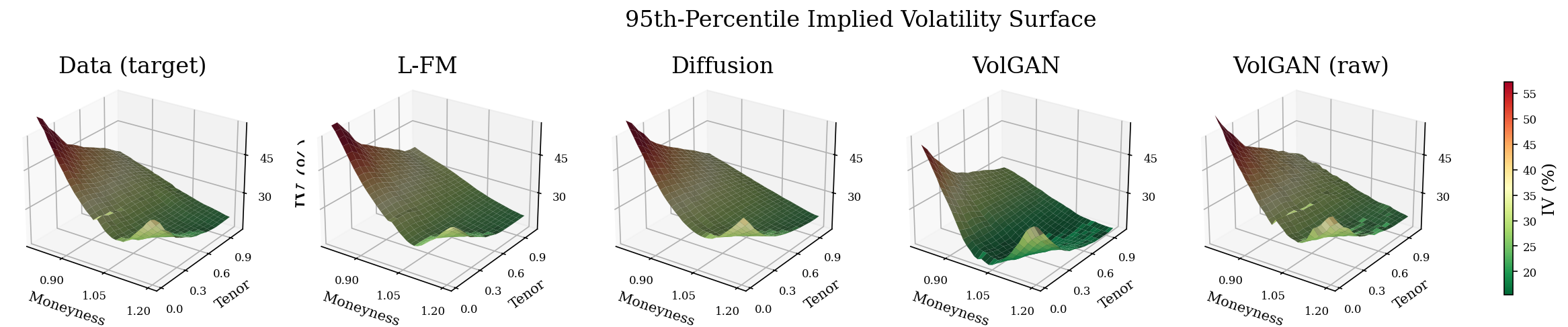}

    \includegraphics[width=0.99\textwidth]{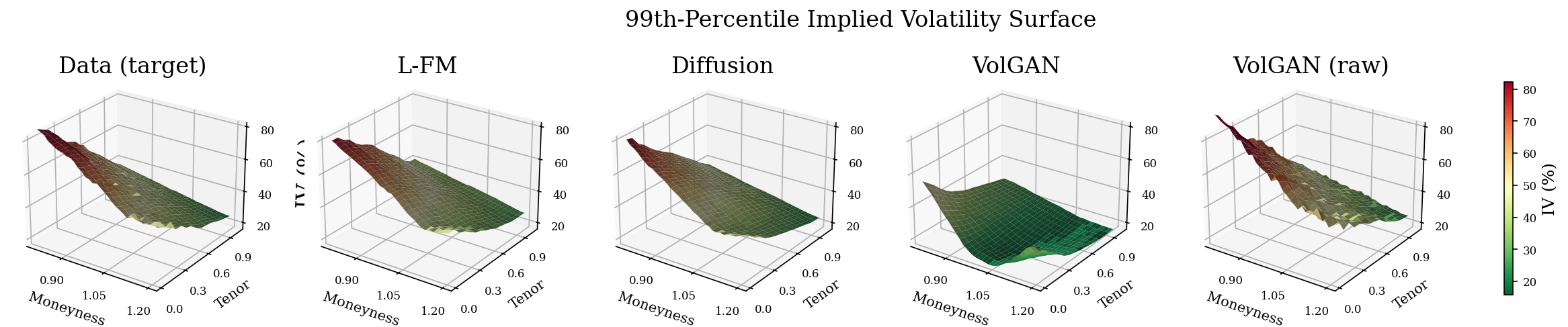}

    \caption{Pointwise $5$th-, $95$th-, and $99$th-percentile
    IVS for the empirical data and the main
    generative models.}
    \label{fig:quantile_surfaces}
\end{figure*}

Tail behavior is evaluated through pointwise quantile surfaces. For a quantile level $q$, each grid value is obtained by taking the empirical $q$-quantile across all surfaces at the corresponding maturity--moneyness point. Figure~\ref{fig:quantile_surfaces} compares the $5$th-, $95$th-, and $99$th-percentile surfaces. At the $5$th percentile, all main models reproduce the overall empirical geometry reasonably well. Larger differences arise in the upper tail. At the $95$th and $99$th percentiles, L-FM and Diffusion preserve the elevated short-maturity put wing and the overall moneyness--maturity structure of the empirical surfaces.  VolGAN produces substantially lower and smoother upper-quantile surfaces, indicating an underrepresentation of extreme volatility levels. VolGAN Raw remains close to the empirical upper-tail surfaces but exhibits more pronounced local irregularities.

Table~\ref{tab:w1_quantiles} shows that VolGAN Raw best matches the empirical quantiles from \(Q_5\) to \(Q_{95}\), while L-FM  ranks second over most quantiles and achieves the lowest error at \(Q_{99}\), whereas Diffusion and VolGAN deteriorate substantially at high quantiles.

\begin{table}[ht]
\centering
\caption{$\mathcal{W}_1$ distance in implied-volatility percentage points
between generated and empirical pointwise quantile surfaces
(mean $\pm$ std). Lower is better.}
%\caption{$\mathcal{W}_1$ distance between generated and empirical
%pointwise quantile surfaces (mean $\pm$ std). Lower is better.}
\label{tab:w1_quantiles}
%\small
% \setlength{\tabcolsep}{6pt}
% \resizebox{\linewidth}{!}{%
\begin{tabular}{lcccc}
\toprule
Quant.
& L-FM
& Diffusion
& VolGAN
& VolGAN Raw \\
\midrule
$Q_{5}$
& $0.40 \pm 0.11$
& $0.44 \pm 0.07$
& \underline{$0.32 \pm 0.19$}
& \textbf{0.24 $\pm$ 0.06} \\
$Q_{25}$
& \underline{0.32 $\pm$ 0.08}
& $0.53 \pm 0.05$
& $1.08 \pm 0.59$
& \textbf{0.22 $\pm$ 0.07} \\
$Q_{50}$
& \underline{$0.35 \pm 0.07$}
& $0.47 \pm 0.06$
& $2.50 \pm 0.72$
& \textbf{0.28 $\pm$ 0.07} \\
$Q_{75}$
& \underline{$0.32 \pm 0.09$}
& $0.64 \pm 0.22$
& $4.60 \pm 0.69$
& \textbf{0.25 $\pm $0.18} \\
$Q_{95}$
& \underline{$0.95 \pm 0.72$}
& $1.55 \pm 0.74$
& $8.84 \pm 0.74$
& \textbf{0.62 $\pm$ 0.39} \\
$Q_{99}$
& \textbf{1.56 $\pm$ 0.71}
& $8.72 \pm 4.36$
& $22.99 \pm 1.16$
& \underline{$2.57 \pm 1.43$} \\
\bottomrule
\end{tabular}%}
\end{table}

\subsubsection{Financial Metrics Diagnostics.} For each factor in \eqref{eq:def-financial-factors}, we compute the \(\mathcal{W}_1\) distance between its empirical distributions over the training and generated surfaces.
Table~\ref{tab:w1_factors} shows that L-FM ranks first for level and curvature and second for skew and term structure. %shows that L-FM achieves the best result for level and curvature, and the second-best results for skew and term structure, yielding a balanced overall performance. 

\begin{comment}
\begin{table}[ht]
\centering
\caption{$\mathcal{W}_1$ distance between model and empirical distributions of financial factors \eqref{eq:def-financial-factors}. Lower is better.}
\label{tab:w1_factors}
\resizebox{\linewidth}{!}{%
\begin{tabular}{lcccc}
\toprule
Factor & L-FM & Diffusion & VolGAN & VolGAN Raw \\
\midrule
Level          & \textbf{0.00388} & 0.00848 & 0.04556 & \underline{0.00462} \\
Skew           & \underline{0.00328} & 0.00929 & 0.01643 & \textbf{0.00316} \\
Term Structure & \underline{0.00859}          & 0.00980 & 0.02611 & \textbf{0.00579} \\
Curvature      & \textbf{0.00263}          & 0.00338 & 0.01620 & \underline{0.00286} \\
\bottomrule
\end{tabular}
}
\end{table} \Oscar{J'ai seulement mis les moyenne ici}
\Yating{vous pourrez me donner les std, je vais essayer d'ajouter}
\end{comment}

\begin{table}[ht]
\centering
\caption{$\mathcal{W}_1$ distance between model and empirical distributions of financial factors \eqref{eq:def-financial-factors}. Lower is better.}
\label{tab:w1_factors}
\resizebox{\linewidth}{!}{%
\begin{tabular}{lcccc}
\toprule
Factor & L-FM & Diffusion & VolGAN & VolGAN Raw \\
\midrule
Level          & \textbf{0.00388} & 0.00848 & 0.04556 & \underline{0.00462} \\
& $\pm$ 0.00049 & $\pm$ 0.00337 & $\pm$ 0.00714 & $\pm$ 0.00161\\
Skew           & \underline{0.00328} & 0.00929 & 0.01643 & \textbf{0.00316} \\
& $\pm$ 0.00202 & $\pm$ 0.00276 & $\pm$ 0.00228 & $\pm$ 0.00100 \\
Term Struc. & \underline{0.00859}          & 0.00980 & 0.02611 & \textbf{0.00579} \\
&$\pm$ 0.00119 & $\pm$ 0.00109 & $\pm$ 0.00445 & $\pm$ 0.00267\\ 
Curvature      & \textbf{0.00263}          & 0.00338 & 0.01620 & \underline{0.00286} \\
&$\pm$ 0.00060 & $\pm$ 0.00097 & $\pm$ 0.00324 & $\pm$ 0.00147 \\
\bottomrule
\end{tabular}
}
\end{table}

\subsubsection{No-Arbitrage Validity.}
Table~\ref{tab:zero_arb} reports the percentage of surfaces satisfying the calendar-spread, call-spread, and butterfly conditions over the entire grid. L-FM achieves the highest validity rate, \(90.8\%\), followed by VolGAN at \(69.7\%\), while Diffusion and VolGAN Raw attain \(46.7\%\) and \(32.1\%\), respectively. Only \(51.2\%\) of the empirical surfaces satisfy all three conditions, possibly due to market noise, numerical inversion errors, and interpolation artifacts. The arbitrage penalties therefore regularize L-FM toward smoother and more financially consistent surfaces.

%Table~\ref{tab:zero_arb} reports the percentage of surfaces satisfying the calendar-spread, call-spread, and butterfly conditions over the entire grid. VolGAN  achieves the highest validity rate, $92.3\%$, closely followed by L-FM at $90.8\%$. These values are substantially higher than those obtained by Diffusion ($58.7\%$) and VolGAN Raw ($39.4\%$). It is worth noting that only $51.2\%$ of the training surfaces satisfy all three static no-arbitrage conditions, possibly due to market noise, numerical inversion errors, and interpolation artifacts. The arbitrage penalties therefore regularize the model toward smoother and more financially consistent surfaces.

\begin{table}[ht]
\centering
\caption{Percentage of surfaces satisfying all three static no-arbitrage conditions over the entire grid. Higher is better.}
\label{tab:zero_arb}
\resizebox{\linewidth}{!}{%
\begin{tabular}{ccccc}
\toprule
Data & L-FM   & Diffusion    & VolGAN  & VolGAN Raw   \\
\midrule
76.6\% & \textbf{90.8}\% $\pm 1.5\%$ & $46.7\%\pm 4.3\%$ &$\underline{69.7}\%\pm 8.8\%$&$32.1\%\pm 3.0\%$\\
\bottomrule
\end{tabular}}
\end{table}

\subsubsection{Conclusion} 
Taken together, the distributional, shape, tail, and no-arbitrage results show that L-FM provides a favorable compromise. It closely matches the empirical distribution, reproduces the main smile and skew structures, performs best in the extreme $Q_{99}$ regime, and achieves a $90.8\%$ no-arbitrage validity rate. It therefore remains competitive with VolGAN Raw in distributional similarity while improving extreme-tail reproduction and increasing the proportion of arbitrage-free surfaces from 32.1\% to 90.8\% (Table \ref{tab:zero_arb}), without requiring the post-sampling reweighting used in VolGAN. %It therefore remains close to VolGAN Raw in distributional similarity, better preserves the extreme upper tail than the other regularized baselines, and approaches the financial validity of VolGAN, without requiring post-sampling reweighting.

\subsection{Ablation and Sensitivity Analyses}\label{subsec:Ablation}

We now examine the VAE loss design, the arbitrage--fidelity trade-off, the latent-space dimension, and the contribution of the  flow matching component.

\subsubsection{Effect of the VAE Loss Design}

Table~\ref{tab:ablation} examines the contribution of the VAE loss components. Setting \(\beta=0\) moderately increases the Global \(\mathcal{W}_1\) distance and the SWD, while slightly reducing the arbitrage-free rate. Removing the calendar penalty substantially reduces the proportion of arbitrage-free surfaces and increases the Global \(\mathcal{W}_1\), despite a slightly lower SWD. Removing the butterfly penalty also deteriorates the distributional metrics and lowers the arbitrage-free rate, supporting the role of both arbitrage penalties in controlling surface validity. Since no call-spread arbitrage is observed in the training data, we set \(\lambda_{\mathrm{call}}=0\) in the full model. Introducing a positive call-spread penalty %slightly improves the arbitrage-free rate but 
worsens both distributional metrics but leaves the arbitrage-free rate unchanged, indicating that it provides no clear overall benefit for the present dataset.

\begin{table}[ht]
\centering
\caption{Ablation study of the VAE loss components. Results are reported as mean $\pm$ standard deviation, with \% arb-free denoting the proportion of arbitrage-free generated surfaces.}
\label{tab:ablation}
\resizebox{\linewidth}{!}{%
\begin{tabular}{lccccc}
\toprule
Variant
  & {Global $\mathcal{W}_1$}
  & {SWD}
  & \% arb-free \\
%\cmidrule(lr){2-3}\cmidrule(lr){4-5}
%  & Mean (Std) & Mean (Std)
%  & \% arb-free \\
\midrule
Full model 
& $0.0028 \pm 0.0014$ & $0.0069 \pm 0.0005$
  & $90.8\% \pm 1.5\%$\\
\midrule
Set $\beta=0$
& $0.0037 \pm 0.0008$
& $0.0078 \pm 0.0006$
  & $89.8\% \pm 0.4\%$  \\
Set  $\lambda_{\mathrm{cal}}=0$
& $0.0054 \pm 0.0051$
& $0.0067 \pm 0.0036$
  & $78.6\% \pm 3.6\%$ \\
Set $\lambda_{\mathrm{but}}=0$
& $0.0055 \pm 0.0044$
& $0.0090 \pm 0.0023$
& $84.3\% \pm 6.4\%$  \\
Set $\lambda_{\mathrm{call}}\!=\!0.03$
& $0.0033 \pm 0.0011$
& $0.0079 \pm 0.0002$
  & $90.8\% \pm 1.5\%$  \\
\bottomrule
\end{tabular}
}
\end{table}

\subsubsection{Distributional Cost of Nearly Arbitrage-Free Generation} To quantify the cost of stronger arbitrage control, we increase the penalty coefficient and retain the smallest value achieving 99.9\% arbitrage-free generation. Table~\ref{tab:lfm_results} shows a marked deterioration in both global metrics and pointwise quantile accuracy, with the largest discrepancies occurring in the upper tail. This highlights a clear trade-off between near-complete arbitrage elimination and distributional fidelity, particularly for extreme volatility regimes.

\begin{table}[ht]
\centering
\caption{Global (top) and quantile-wise (bottom) distances for L-FM achieving 99.9\% arbitrage-free generation.
}
\label{tab:lfm_results}
\begin{tabular}{lcc}
\toprule
Metric & Global $\mathcal{W}_1$ & SWD \\
\midrule
L-FM &  0.03058  & 0.04753 \\
\bottomrule
\end{tabular}

\smallskip

\begin{tabular}{lcccccc}
\toprule
Quantile & $Q_5$&  $Q_{25}$ &  $Q_{50}$ &   $Q_{75}$ &  $Q_{95}$ &  $Q_{99}$\\
\midrule
L-FM & 1.986 & 0.864 & 2.407 & 4.742 & 8.951 & 23.778 \\
\bottomrule
\end{tabular}
\end{table}

\subsubsection{Sensitivity to the Latent-Space Dimension}

Table~\ref{tab:zdim} evaluates the sensitivity of L-FM to the latent-space dimension using global $\mathcal{W}_1$ distances and SWD, upper-tail quantile errors, and the arbitrage-free rate, using 1,000  epochs for the VAE and 400 epochs for flow matching. Performance is broadly stable across dimensions, while \(d=6\) provides the best overall trade-off, achieving the lowest Global  $\mathcal{W}_1$, SWD, and $Q_{99}$ error with a comparable arbitrage-free rate.

\begin{table}[ht]
\centering
\caption{Sensitivity to the Latent-Space Dimension}
\label{tab:zdim}
\begin{tabular}{lcccc}
\toprule
Dim.
  & Global $\mathcal{W}_1$ & SWD
  & \% arb-free \\
\midrule
4
  & $0.00400 \pm 0.00109$ & $0.00798 \pm 0.00069$
  & $89.3 \pm 1.5\%$ \\
6
  & $0.00272 \pm 0.00117$ & $0.00769 \pm 0.00082$
  & $89.3 \pm 2.4\%$ \\
8
  & $0.00316 \pm 0.00115$ & $0.00788 \pm 0.00035$
  & $88.9 \pm 0.9\%$ \\
10
  & $0.00321 \pm 0.00065$ & $0.00793 \pm 0.00016$
  & $89.2 \pm 2.2\%$ \\
\bottomrule
\end{tabular}

\smallskip

\begin{tabular}{lccc}
\toprule
Dim.
  & $\mathcal{W}_1$ - $Q_{95}$ & $\mathcal{W}_1$ - $Q_{99}$
 \\
\midrule
4
  & $0.65523 \pm 0.18971$ & $5.05401 \pm 3.41251$ \\
6
  & $1.30459 \pm 0.76485$ & $1.77397 \pm 1.43587$\\
8
  & $2.34132 \pm 0.21715$ & $3.55578 \pm 0.89553$ \\
10
  & $1.90629 \pm 0.86982$ & $3.71988 \pm 2.04751$\\
%12
%  & $1.70095 \pm 0.88230$ & $5.58339 \pm 1.39282$\\
\bottomrule
\end{tabular} 
\end{table}

\subsubsection{Contribution of Flow Matching Component}

To assess the contribution of the flow matching component, we consider an ablation study in which samples are generated using only the VAE, without the  flow matching step. Table~\ref{tab:vae-only} reports the corresponding results, and the VAE-only model achieves an arbitrage-free rate of $88.98 \pm 1.65\%$, which is close to that of the L-FM model reported in Table~\ref{tab:zero_arb}.

\begin{table}[ht]
\centering
\caption{Results for the VAE-only model (mean $\pm$ std).
}
\label{tab:vae-only}
\begin{tabular}{lcc}
\toprule
Metric & Global $\mathcal{W}_1$ & SWD \\
\midrule
VAE only &  0.00388 ± 0.00072  & 0.00767 ± 0.00039 \\
\bottomrule
\end{tabular}

 \medskip

\begin{tabular}{lc}
\toprule
Quant.
& VAE only  \\
\midrule
$Q_{5}$
& $0.39477 \pm 0.07093$\\
$Q_{25}$
&  $0.26836 \pm 0.06469$ \\
$Q_{50}$
& $0.27128 \pm 0.03394$\\
$Q_{75}$
& $0.30855 \pm 0.06888$\\
$Q_{95}$
& $1.59484 \pm 0.26341$\\
$Q_{99}$
& $5.80845 \pm 0.55807$\\
\bottomrule
\end{tabular}%}
\end{table}

Compared with the L-FM results reported in Tables \ref{tab:wasserstein} and \ref{tab:w1_quantiles},  the VAE-only model performs worse in terms of global distributional accuracy, but slightly better at the lower and central quantiles ($Q_5$ to $Q_{75}$). However, L-FM provides a clear improvement in the accuracy of the upper quantiles, reducing the errors at $Q_95$ and $Q_99$ by approximately 40\% and 73\% respectively. Similar limitations of VAEs in capturing extreme values have been observed in the literature (see, e.g., \cite{ziani2026phase}). The latent flow matching component therefore appears beneficial for capturing the upper-tail behavior of implied volatility surfaces.

Finally, we emphasize that the arbitrage-aware loss is incorporated into the VAE training stage. Consequently, using only a flow matching model would remove this explicit arbitrage-aware mechanism from our current framework.

% =====================================================
% 5. Limitations and Conclusion
% =====================================================

\section{Conclusion and Discussion}
\label{sec:conclusion}

We proposed an arbitrage-aware latent flow-matching (L-FM) framework for implied volatility surface generation. The method combines an arbitrage-regularized VAE with a latent flow-matching model, thereby reducing the dimensionality of the generation problem while incorporating financial structure into the learned representation and decoder.

Empirical results show that L-FM achieves a favorable balance between distributional similarity, tail preservation, and financial consistency. It closely matches the empirical implied volatility distribution, reproduces the main smile and skew structures, performs
best in the extreme $Q_{99}$ regime, and achieves a $90.8\%$ no-arbitrage validity rate. L-FM therefore remains close to VolGAN Raw in distributional distance, better captures the extreme upper tail, and achieves the highest no-arbitrage validity rate without requiring post-sampling reweighting. %L-FM therefore remains close to VolGAN Raw in distributional distance, better preserves the extreme upper tail than the regularized baselines, and approaches the validity of VolGAN without requiring post-sampling reweighting.

The present study nevertheless has several limitations. First, static no-arbitrage is encouraged through soft penalties rather than enforced by construction, so validity is assessed empirically and is not guaranteed for every generated surface. Second, the model is trained and evaluated on a fixed maturity--moneyness grid, and its behavior outside this grid is not examined. Finally, constrained by the limited amount of available data, the current framework is unconditional and is designed to reconstruct and sample from the empirical distribution of implied volatility surfaces, rather than to forecast their future evolution.  Forecasting and conditional generation would additionally require information on the current market state and temporal dynamics.

Despite these limitations, the proposed framework offers a promising foundation for synthetic scenario generation, stress testing, risk analysis, and data augmentation. Future work will extend the model to conditional generation using market variables and to forecasting the temporal evolution of implied volatility surfaces, while also investigating stronger arbitrage-preserving parameterizations.

% =====================================================
% Acknowledgments
% =====================================================
\begin{acks}
This project has received financial support from the CNRS through the MITI interdisciplinary programs. Yating Liu gratefully acknowledges Grégoire Szymanski for valuable discussions related to this
work.
\end{acks}

% =====================================================
% References
\bibliographystyle{ACM-Reference-Format}
\bibliography{main-bib}

\end{document}